\documentclass[12pt]{iopart}

\usepackage{amssymb}
\usepackage{iopams}
\newcommand{\eqref}[1]{(\ref{#1})}
\usepackage{rotating}
\usepackage{color}
\usepackage{cancel}
\usepackage{orcidlink}
\usepackage{doi}
\begin{document}
\title[Path Integral Formulation of Wigner-Dunkl Quantum Mechanics]{On the Path Integral Formulation of Wigner-Dunkl Quantum Mechanics}
\author{Georg Junker
\orcidlink{0000-0003-2054-0453}}
\address{Institut für Theoretische Physik I, Friedrich-Alexander-Universität, Staudtstr. 7, 91058 Erlangen, Germany\\[2mm] and \\[2mm]}
\address{European Southern Observatory, Karl-Schwarzschild-Stra{\ss}e 2, 85748 Garching, Germany}

\eads{\mailto{georg.junker@fau.de}, \mailto{gjunker@eso.org}}
\begin{abstract}
Feynman's path integral approach is studied in the framework of the Wigner-Dunkl deformation of quantum mechanics. We start with reviewing some basics from Dunkl theory and investigate the time evolution of a Gaussian wave packet, which exhibits the same dispersion relation as observed in standard quantum mechanics. Feynman's path integral approach is then extended to Wigner-Dunkl quantum mechanics. The harmonic oscillator problem is solved explicitly. We then look at the Euclidean time evolution and the related Dunkl process. This process, which exhibit jumps, can be represented by two continuous Bessel processes, one with reflection and one with absorption at the origin. The Feynman-Kac path integral for the harmonic oscillator problem is explicitly calculated.
\end{abstract}

\noindent{\it Keywords\/}: Wigner-Dunkl quantum mechanics, Feynman path integrals, Feynman-Kac formula, Stochastic processes
%\maketitle

\section{Introduction}
Since the early 1990s deformed quantum mechanics has attracted much attention in the mathematical and theoretical physics community. The so-called $q$-deformed quantum mechanics is based on quantum groups, which in essence are deformations of the Heisenberg algebra. It goes at least back to the work by Arik and Coon \cite{Arik1976} who studied the $q$-deformed harmonic oscillator algebra $aa^\dag = q a^\dag a +1$ with $0<q<1$ and associated coherent states. See also ref.\ \cite{Kuryshkin1980}. This $q$-deformed quantum mechanics became popular with the seminal work by Macfarlane \cite{Macfarlane1989} and Biedenharn \cite{Biedenharn1989} and is still a topic of intensive research. %See for example \cite{Dobrev2017}.

In 1950 Eugene Wigner \cite{Wigner1950}, when investigating the question "Do the Equations of Motion Determine the Quantum Mechanical Commutation Relations?", found that these relations are not necessarily unique. As shown by Yang \cite{Yang1951} a year later, Wigner's finding can be seen as another form of a deformed harmonic oscillator algebra involving the reflection operator $R$ acting on position eigenstates as follows $R|x\rangle=-|x\rangle$. More explicitly, Yang proposed for the $x$-representation of the momentum operator a deformed derivative given by $\tilde{\partial}:=\partial_x +(c/2x)R$ with arbitrary constant $c$ being the deformation parameter. Soon after, this work by Wigner and Yang stimulated Green \cite{Green1953} to propose a generalisation of field quantisation beyond those for bosons and fermions resulting in concepts like parafields and parastatistics \cite{Volkov1959,Volkov1960}. For subsequent work related to fraction spin fields, supersymmetric quantum mechanics, Bose-Fermi transformations, parabosons, parafermions and other topics we refer to the work by Plyushchay and reference therein \cite{Plyushchay1994,Plyushchay1994b,Plyushchay1996,Plyushchay1996b,Plyushchay1997,Plyushchay2000}.

Independently of the above work, Charles Dunkl \cite{Dunkl1989} studied differential-difference operators related to reflection groups, which for the simplest reflection group $Z_2$ acting on the real line $x\in\mathbb{R}$, may be represented by the Dunkl derivative
\begin{equation}\label{Dx}
  D_x:= \frac{\partial}{\partial x} +\frac{\nu}{x}\left(1-R\right)\,.
\end{equation}
As in the case of Yang, the reflection operator $R$ acts on functions $f:\mathbb{R}\mapsto\mathbb{C}$ defined on the real line as follows:
\begin{equation}\label{Rdef}
  (Rf)(x) := f(-x)\,.
\end{equation}
Indeed, soon later, the close connection between Wigner's work and that of Dunkl has been realised and pointed out by Rosenblum \cite{Rosenblum1994} in his studies of the Wigner-Dunkl harmonic oscillator. The deformed Heisenberg algebra providing the basis of all these investigations is given by
\begin{equation}\label{commute}
  [P,x] = \frac{\hbar}{\rmi}\left(1 +2\nu R\right)
\end{equation}
with the Dunkl momentum operator being represented by the Dunkl derivative \eqref{Dx} as
\begin{equation}\label{P}
  P:=\frac{\hbar}{\rmi} D_x\,.
\end{equation}
This momentum operator is a self-adjoint operator, that is,
\begin{equation}\label{selfadj}
  \langle \phi|P \psi\rangle = \langle P\phi|\psi\rangle \qquad{\rm or}\qquad P^\dag = P\,,
\end{equation}
on the weighted Hilbert space ${\cal H}(\mathbb{R},\rmd x\,|x|^{2\nu})$ where the scalar product is given
\begin{equation}\label{prod}
\langle \phi| \psi\rangle := \int_{-\infty}^{+\infty}\rmd x \, |x|^{2\nu}\phi^*(x)\psi(x)\,,\qquad \nu > - \frac{1}{2}\,.
\end{equation}
Here we note that the deformation parameter $\nu$ is bounded from below in order to result in a well-defined measure.

Quantum physics based on this setup is called Wigner-Dunkl quantum mechanics. Within this framework completeness and orthogonality of position eigenstates read
\begin{equation}\label{unitop&ortho}
  1=  \int_{-\infty}^{+\infty}\rmd x \,|x|^{2\nu} |x\rangle\langle x|\,,\qquad\langle y |x\rangle = \frac{1}{|x|^{2\nu}}\,\delta(x-y)\,.
\end{equation}
Furthermore, a generic non-relativistic Wigner-Dunkl Hamiltonian for a scalar potential $V:\mathbb{R}\mapsto\mathbb{R}$ is then given
\begin{equation}\label{H}
  H := \frac{P^2}{2m} +V(x) = -\frac{\hbar^2}{2m}\left[\frac{\partial^2}{\partial x^2} +\frac{2\nu}{x}\frac{\partial}{\partial x}-\frac{\nu}{x^2}(1-R)\right] +V(x)\,.
\end{equation}
Such Wigner-Dunkl quantum systems had been studied extensively during the last two decades and are still attracting much attention in the current literature. We can provide here only a few, more recent, references and refer to further literature given therein.  For example, a discussion of the free particle, the particle in a box, the harmonic oscillator and Coulomb problem see \cite{Rodridues2009,Horvathy2010,Genest2013,Genest2014,Genest2014b,Chung2019,Ghazouani2021,Mota2022,Dong2023}. The associated coherent states are studied in \cite{Ghazouani2022,Sedaghatnia2023}. The coupling to electromagnetic fields was discussed in \cite{Chung2021,Junker2023}. For relativistic aspects see \cite{Mota2019,Hassanabadi2022} and references therein.

The purpose of the present paper is to investigate the path integral formulation of Wigner-Dunkl quantum mechanics (WDQM). It appears that till today only the Schrödinger, Klein-Gordon and Dirac equations of WDQM had been discussed in the literature. The path integral approach has not yet been considered. Here we will follow Feynman's \cite{Feynman1948,Feynman2010,Schulman2012} idea to formulate a path integral approach to WDQM. In doing so we will start collecting some important findings on the Dunkl theory from the mathematical literature in the next section. Equipped with these tools, in section 3, we take a closer look at the Wigner-Dunkl version of Schrödinger's equation in one dimension and explicitly discuss the time evolution of a Gaussian wave packet on the real line.
In Section 4 we then construct the propagator, that is, the time evolution operator in $x$-representation, for the free particle in WDQM. With the help of the Lie-Trotter product formula we can then establish a path integral formulation of Wigner-Dunkl quantum mechanics, which is discussed in some details. Path integration is explicitly performed for the harmonic oscillator problem in WDQM.
Section 5 is dedicated to the Euclidean time evolution. We take a closer look at the Dunkl process and its related Feynman-Kac formula. It is shown that the Dunkl process on the real line exhibiting jumps can be reduced to two Bessel processes with continuous paths on the positive half-line. The Feynman-Kac path integral for the harmonic oscillator is explicitly calculated.
We conclude in section 6 with a brief summary and an outlook for further investigations. For the readers convenience a few basic results on Bessel processes are collected in Appendix A.

\section{Preliminaries on Dunkl kernel and transform}
In this section we will summarise several findings known from the Dunkl theory. Hence, this section does not contain any new results but presents known facts in a way most suitable for our way forward
\cite{Dunkl1991,Jeu1993,Rosenblum1994,Roesler1998a,Roesler1999}. Let us start by introducing the Dunkl number as a deformation of the natural numbers,
\begin{equation}\label{Dunkln}
  [n]_\nu := n + \nu[1-(-1)^n]\,,\qquad n\in\mathbb{N}_0\,.
\end{equation}
Obviously, this deformation has an effect on odd numbers only. More explicitly, we have
\begin{equation}\label{0123}
  [2m]_\nu = 2m\,,\qquad [2m+1]_\nu = 2m+1+2\nu\,,\qquad m\in\mathbb{N}_0\,.
\end{equation}
Furthermore, we introduce the Dunkl factorial \cite{Rosenblum1994}
\begin{equation}\label{Dunkln!}
  [n]_\nu !:= [1]_\nu [2]_\nu [3]_\nu \cdots [n]_\nu\,,\qquad [0]_\nu ! := 1\,,
\end{equation}
and note that this leads us to explicit expressions of the form
\begin{equation}\label{Dunkln!2}
\fl
 \textstyle
  [2m]_\nu != 2^{2m}m!(\nu+\frac{1}{2})_m \,,\qquad [2m+1]_\nu != 2^{2m+1}m!(\nu+\frac{1}{2})(\nu+\frac{3}{2})_m\,,
\end{equation}
with the Pochhammer symbol being defined via the Gamma function by
$$
(z)_n:= \frac{\Gamma(z+n)}{\Gamma(z)}=z(z+1)(z+2)\cdots(z+n-1).
$$
\subsection{The Dunkl kernel}
With the help of the Dunkl factorial we may now define a deformed exponential function as follows \cite{Rosenblum1994}
\begin{equation}\label{E_v}
 \begin{array}{rl}
  E_\nu(z) &:= \displaystyle\sum_{n=0}^{\infty}\frac{z^n}{[n]_\nu!}\\
           & = \displaystyle\sum_{m=0}^{\infty}\frac{(z/2)^{2m}}{m!(\nu+\frac{1}{2})_m} + \frac{z}{2\nu+1} \sum_{m=0}^{\infty}\frac{(z/2)^{2m}}{m!(\nu+\frac{3}{2})_m}\\[4mm]
           & = \,_0F_1(\nu +\frac{1}{2},\frac{z^2}{4}) + \frac{z}{2\nu+1} \,_0F_1(\nu +\frac{3}{2},\frac{z^2}{4})
\end{array}
\end{equation}
This deformed exponential function is also called Dunkl kernel \cite{Dunkl1991} due to its importance in the so-called Dunkl transformations \cite{Jeu1993}. Being more precise, \eqref{E_v} is the Dunkl kernel related to the reflection group $Z_2$ in one dimension generated by $R$.
Obviously, the Dunkl kernel is an entire function consisting of an even and an odd term as indicated above. Here $_0 F_1(c,z)$ denotes a generalised hypergeometric function. Alternative expressions in terms of the  confluent hypergeometric function are
\begin{equation}\label{E_v2}
  E_\nu(z) = \rme^z\,_1F_1(\nu,2\nu+1,-2z)  = \rme^{-z}\,_1F_1(\nu+1,2\nu+1,2z)
\end{equation}
For real argument $x\in\mathbb{R}$ we may also use an expression in terms of the modified Bessel function of the first kind denoted by ${\rm I}_\alpha (x)$:
\begin{equation}\label{Ereal}
  E_\nu(x) = \textstyle\Gamma(\nu+\frac{1}{2})\displaystyle \left(\frac{2}{|x|}\right)^{\nu-\frac{1}{2}}\left[{\rm I}_{\nu-\frac{1}{2}}(|x|)+ {\rm sgn\,}x\,{\rm I}_{\nu+\frac{1}{2}}(|x|)\right]\qquad
\end{equation}
For purely imaginary argument $z=\rmi x$ with $x\in\mathbb{R}$ the Dunkl kernel can be represented by the Bessel function of the first kind \cite{Rosenblum1994,Roesler1999,Roesler2002}, which follows from the relation ${\rm J}_{\alpha}(x) = \frac{(x/2)^\alpha}{\Gamma(\alpha +1)}\,_0F_1(\alpha+1,-\frac{x^2}{4})$ resulting in
\begin{equation}\label{Ecompl}
  E_\nu(\rmi x) = \frac{\Gamma(\nu+\frac{1}{2})}{(|x|/2)^{\nu-\frac{1}{2}}}
  \left[{\rm J}_{\nu-\frac{1}{2}}(|x|)+\rmi\,{\rm sgn\,}x\, {\rm J}_{\nu+\frac{1}{2}}(|x|)\right]\,.
\end{equation}
Note that $E_\nu(-\rmi x)=E^*_\nu(\rmi x)$. We also note that $E_\nu(0) = 1$ and for $\nu =0$ we arrive at the undeformed exponential function, i.e., $E_0(z)=\rme^z$.

For large $x>0$ we make use of representation \eqref{E_v2} together with the asymptotic form of the confluent hypergeometric function
$$
\,_1F_1(\nu+1,2\nu+1,2x)=\frac{\rme^{2x}\Gamma(2\nu+1)}{(2x)^\nu\Gamma(\nu+1)}\left(1-\frac{\nu^2}{2x}+O(x^{-2})\right)
$$
to arrive at the asymptotic expressions for $x\to+\infty$
\begin{equation}\label{Elargex}
\begin{array}{l}
  \displaystyle E_\nu(x) = \frac{c_\nu}{\sqrt{2\pi}} \frac{\rme^{x}}{x^\nu}\left(1-\frac{\nu^2}{2x}+O(x^{-2})\right)\,,\\[4mm]
  \displaystyle E_\nu(-x) = \frac{\nu c_\nu}{2\sqrt{2\pi}} \frac{\rme^{x}}{x^{\nu+1}}\left(1-\frac{\nu^2-1}{2x}+O(x^{-2})\right)\,,
\end{array}
\end{equation}
where we have introduced the number
\begin{equation}\label{Cnu}
  c_\nu:=\textstyle 2^{\nu+1/2}\Gamma(\nu+\frac{1}{2}) = \displaystyle\sqrt{2\pi}\,\frac{\Gamma(2\nu+1)}{2^\nu\Gamma(\nu+1)}\,.
\end{equation}

Noting that
\begin{equation}\label{Dxn}
  D_x \, x^n = n\,x^{n-1}+ \nu x^{n-1}(1-(-1)^n) = [n]_\nu x^{n-1}\,, n\in\mathbb{N}_0
\end{equation}
we also conclude that
\begin{equation}\label{E_vP}
(D_x E_\nu)(ax) = aE_\nu(ax)\,,\quad a\in\mathbb{C}\,.
\end{equation}
In particular, we have
\begin{equation}\label{Ppsi}
P\,E_\nu(\rmi k x)= \hbar k\,E_\nu(\rmi k x)\,,\qquad k\in\mathbb{R}\,,
\end{equation}
indicating that the deformed exponential function with complex argument is an eigenfunction of the Dunkl momentum with eigenvalue $\hbar k$.

\subsection{The Dunkl transform}
With the Dunkl kernel being a deformed exponential function one may introduce a deformed Fourier transformation, the so-called Dunkl transformation, being defined by \cite{Jeu1993,Roesler1999}
\begin{equation}\label{Df}
  {\cal D}[f]\equiv g(k):=\displaystyle \frac{1}{c_\nu}\int_{-\infty}^{\infty}\rmd x \, |x|^{2\nu}\,f(x)\,E_\nu(-\rmi kx)
\end{equation}
with inverse transformation given by
\begin{equation}\label{Dinvg}
  {\cal D}^{-1}[g]\equiv f(x):=\displaystyle \frac{1}{c_\nu} \int_{-\infty}^{\infty}\rmd k \, |k|^{2\nu}\,g(k)\,E_\nu(\rmi kx)\,.
\end{equation}
Note that we at least require $f,g\in L^1(\mathbb{R},\rmd x |x|^{2\nu})$.
Obviously for $\nu = 0$ we have $c_0=\sqrt{2\pi}$ and above transformation reduces to the well-known Fourier transformation on the real line. For even functions $(Rf)(x)=f(x)$ the Dunkl transformation in essence is equivalent to Hankel's transformation of order $\nu-1/2$ for the function $F(x) = |x|^{2\nu-1}f(x)/(2c_\nu)$ as only the real part of the Dunkl kernel contributes in the integral \eqref{Df} and reduces it to
$$
{\cal D}[f]=\int_{0}^{\infty}\rmd x\, x\, F(x)\,{\rm J}_{\nu-1/2}(kx)\,,
$$
which indeed is the Hankel transformation for index $\nu-1/2$ \cite{Roesler1999}.

There are two important examples of Dunkl transforms, which we like to mention here. The first one is
\begin{equation}\label{DE}
  {\cal D}[E_\nu(\rmi\kappa x)]=\frac{c_\nu}{|k|^{2\nu}}\delta(k-\kappa)\,.
\end{equation}
More explicitly we have for $k,\kappa\in \mathbb{R}$
\begin{equation}\label{Ortho}
  \int_{-\infty}^{\infty}\rmd z\,|z|^{2\nu} E^*_\nu(\rmi k z) E_\nu(\rmi \kappa z) =
\frac{c^2_\nu}{|k|^{2\nu}}\,\delta(k - \kappa)\,.
\end{equation}
This may easily be verified via the orthogonality relation of Bessel functions
\begin{equation}\label{BesselOrth}
\int_{0}^{\infty}\rmd z\,z {\rm J}_\alpha(k z){\rm J}_\alpha(\kappa z)=\frac{1}{k}\,\delta(k-\kappa)\,.
\end{equation}

Another important Dunkl transformation is that of a Gaussian function, say $f(x)=\exp\{-\alpha x^2/2 \}$ with ${\rm Re\,}\alpha > 0$.
\begin{equation}\label{DGauss}
\begin{array}{rl}
  {\cal D}[\exp\{-\alpha x^2/2 \}]&=\displaystyle \frac{1}{c_\nu}\int_{-\infty}^{\infty}\rmd x \, |x|^{2\nu}\,\rme^{-\alpha x^2/2 }\,E_\nu(-\rmi kx)\\[4mm]
  &= \displaystyle \frac{1}{k^{2\nu+1}} \int_{0}^{\infty}\rmd z z^{\nu+1/2}\rme^{-\frac{\alpha}{2k^2}z^2}{\rm J}_{\nu-1/2}(z)\\[4mm]
  &= \displaystyle\frac{\exp\{-k^2/2\alpha\}}{\alpha^{\nu+1/2}}\,.
\end{array}\end{equation}
Here in the first step we have reduced the Dunkl transformation to the Hankel transformation as the Gaussian is an even function in $x$. In the second step, the explicit integration is performed with the help of formula 6.629/5 in the table \cite{GR}.
The corresponding inverse transformation follows the same route and is given by
\begin{equation}\label{DinvGauss}
{\cal D}^{-1}[\exp\{-x^2/2\beta \}] = \beta^{\nu+1/2}\exp\{-\beta k^2/2\}\,,\qquad {\rm Re \,}\beta > 0\,.
\end{equation}

\section{Time evolution of a Gaussian wave packet in WDQM}
The above summary of the Dunkl theory allows us to study the time evolution of a Gaussian wave packet within the Wigner-Dunkl formalism. For the free particle of mass $m>0$ on the real line the corresponding Hamiltonian is given by
\begin{equation}\label{Hfree}
  H_\nu := \frac{P^2}{2m}= -\frac{\hbar^2 D_x^2}{2m} = -\frac{\hbar^2}{2m}\left[\frac{\partial^2}{\partial x^2} +\frac{2\nu}{x}\frac{\partial}{\partial x}-\frac{\nu}{x^2}(1-R)\right]\,.
\end{equation}
The associate eigenfunctions are those of the Dunkl momentum operator as given in \eqref{Ppsi}, that is,
\begin{equation}\label{Hpsi}
  H_\nu \psi_k = {\cal E}_k\, \psi_k\,,
\end{equation}
with
\begin{equation}\label{Psifree}
  \psi_k(x)= \frac{|k|^\nu}{c_\nu} \, E_\nu(\rmi k x)\,, \quad {\cal E}_k=\frac{\hbar^2 k^2}{2m}\,, \quad k\in\mathbb{R}\,,
\end{equation}
being proper normalised, see \eqref{Ortho},
\begin{equation}\label{norm}
  \int_{-\infty}^{\infty}\rmd x\,|x|^{2\nu} \psi^*_{k_1}(x)\psi_{k_2}(x) = \delta(k_1-k_2)\,.
\end{equation}
They form a complete set obeying the relation
\begin{equation}\label{compl}
  \int_{-\infty}^{\infty}\rmd k\, \psi^*_{k}(y)\psi_{k}(x) = \frac{1}{|x|^{2\nu}}\,\delta(x-y)\,.
\end{equation}
Note that this is the correct delta function on ${\cal H}=L^2(\mathbb{R},\rmd x\,|x|^{2\nu})$.

With the $k$-wave solution \eqref{Psifree} we may now construct superpositions of such waves defined by
\begin{equation}\label{Psi0}
  \Psi(x):= \int_{-\infty}^{\infty}\rmd k\, a(k)|k|^\nu \psi_k(x)\,.
\end{equation}
Here the spectral weighting function $a(k)$ is required to be normalised by
\begin{equation}\label{Anorm}
\int_{-\infty}^{\infty}\rmd k\, |k|^{2\nu}\, |a(k)|^2 = 1
\end{equation}
in order to lead to a normalised wave packet \eqref{Psi0}. It is obvious that the wave packet is nothing but the inverse Dunkl transform of the spectral function $a(k)$:
\begin{equation}\label{Wavepacket}
\Psi(x)= \int_{-\infty}^{\infty}\rmd k\, \frac{|k|^{2\nu}}{c_\nu} \, a(k) E_\nu(\rmi k x) = {\cal D}^{-1}[a(k)]\,.
\end{equation}
For a Gaussian wave packet this spectral function is given by
\begin{equation}\label{Gausspacket}
a(k)= \frac{1}{\sqrt{\Gamma(\nu+1/2)\beta^{\nu+1/2}}}\rme^{-k^2/2\beta}
\end{equation}
resulting in a well normalised Gaussian wave function in ${\cal H}$, cf.\ eq.\ \eqref{DinvGauss},
\begin{equation}\label{Gausspacket2}
\Psi(x)= \sqrt{\frac{\beta^{\nu+1/2}}{\Gamma(\nu+1/2)}} \rme^{-\beta x^2 /2}\,.
\end{equation}
The time evolution of such a Gaussian wave packet is then given by
\begin{equation}\label{Psit}
\begin{array}{rl}
 \Psi(x,t)& := \exp\{-(\rmi/\hbar)H_\nu t\} \Psi(x)\\
 & \displaystyle =\frac{1}{c_\nu\sqrt{\Gamma(\nu+1/2)\beta^{\nu+1/2}}}
 \int_{-\infty}^{+\infty}\rmd k\, \rme^{-k^2/2\beta}|k|^\nu\rme^{-\rmi\hbar t k^2/2m}\psi_{k}(x)\\
 & \displaystyle = \frac{1}{\sqrt{\Gamma(\nu+1/2)\beta^{\nu+1/2}}}{\cal D}^{-1}[\exp\{ -k^2/2\beta(t) \}]\,,
\end{array}
\end{equation}
where we have set
\begin{equation}\label{betat}
  \beta(t) := \frac{\beta}{1+\rmi\hbar \beta t/m}= \beta \frac{1-\rmi\hbar \beta t/m}{1+(\hbar \beta t/m)^2}\,,
\end{equation}
leading us to the time-dependent Gaussian
\begin{equation}\label{Psit2}
  \Psi(x,t)=\sqrt{\frac{\beta^{\nu+1/2}}{\Gamma(\nu+1/2)[1+\rmi\hbar\beta t/m]^{2\nu+1}}}\exp\{-\beta(t)x^2/2\}\,.
\end{equation}
As a result we find that the quantum dispersion of a Gaussian wave packet in WDQM is identical to that in standard quantum mechanics
\begin{equation}\label{Psi2}
  \fl
  |\Psi(x,t)|^2= \frac{1}{\Gamma(\nu+1/2)}\left(\frac{\beta}{1+\hbar^2\beta^2 t^2/m^2}\right)^{\nu+1/2}\exp\left\{- \frac{\beta x^2}{1+\hbar^2\beta^2 t^2/m^2}\right\}\,.
\end{equation}
Finally let us note that the expectation values for the variance of position and momentum for this distribution are given by
\begin{equation}\label{Deltaxk}
  (\Delta x)^2= \frac{1+\hbar^2\beta^2 t^2/m^2}{2\beta} \,,\qquad (\Delta k)^2 = \frac{\beta}{2}\,,
\end{equation}
resulting in an uncertainty relation for such a Gaussian wave packet being the same as in standard quantum mechanics
\begin{equation}\label{Uncertain}
  (\Delta x)^2(\Delta P)^2= \frac{\hbar^2}{4}\left[1+(\hbar \beta t/m)^2\right]\,,
\end{equation}
which is the minimal uncertainty that can be achieved \cite{Roesler1999}.
\section{Feynman Path Integral for Wigner-Dunkl Quantum Mechanics}
With the results of the previous section we are now in a position to present the path integral approach for WDQM. Feynman's path integral expresses the propagator (or transition amplitude) in terms of a sum over histories \cite{Feynman1948,Feynman2010,Schulman2012}.

Let us start with the free-particle propagator, which is the matrix element of the time evolution operator $\rme^{-(\rmi/\hbar)H_\nu t}$ in $x$-representation,
\begin{equation}\label{Kdef}
  K_\nu(x,y;t):= \langle x|\rme^{-(\rmi/\hbar)H_\nu t}|y\rangle\,,\qquad t\geq 0\,.
\end{equation}
This propagator obeys the Kolmogorov-Chapman relation 
\begin{equation}\label{KCrelation}
  K_\nu(x,y;t_1+t_2)=\int_{-\infty}^{\infty}\rmd z\,|z|^{2\nu}\,K_\nu(x,z;t_2) K_\nu(z,y;t_1)
\end{equation}
and the normalisation condition
\begin{equation}\label{NormK}
  \lim_{t\to 0}K_\nu(x,y;t)= \frac{1}{|x|^{2\nu}}\,\delta(x - y)\,.
\end{equation}
Using its spectral representation
\begin{equation}\label{Kspectral}
\begin{array}{rl}
    K_\nu(x,y;t) & = \displaystyle \int_{-\infty}^{\infty}\rmd k \, \rme^{-\rmi \hbar t k^2/2m}\psi^*_{k}(x)\psi_{k}(y)\\
    & = \displaystyle \frac{1}{c_\nu^2}\int_{-\infty}^{\infty}\rmd k \,|k|^{2\nu}\rme^{-\rmi \hbar t k^2/2m}E_\nu(\rmi k y)E^*_\nu(\rmi k x)\,
\end{array}
\end{equation}
we are able to find a closed-form expression. For this we note that in the above integral only the part being symmetric in $k$ in the product $E_\nu(\rmi k y)E^*_\nu(\rmi k x)$ contributes. Namely
\begin{equation}\label{Kspectral2}
\fl
\begin{array}{rcl}
    K_\nu(x,y;t)  & = & \displaystyle \frac{1}{c_\nu^2}\int_{-\infty}^{\infty}\rmd k \,|k|^{2\nu}\rme^{-\rmi \hbar t k^2/2m}\frac{\Gamma^2(\nu+\frac{1}{2}))}{(|xyk^2|/4)^{\nu-\frac{1}{2}}}\\[2mm]
    &&\times\left[{\rm J}_{\nu-\frac{1}{2}}(|kx|){\rm J}_{\nu-\frac{1}{2}}(|ky|)+ {\rm sgn\,}(x y){\rm J}_{\nu+\frac{1}{2}}(|kx|){\rm J}_{\nu+\frac{1}{2}}(|ky|)\right]\\[4mm]
    & = & \displaystyle \frac{4^{\nu}\Gamma^2(\nu+\frac{1}{2})}{c_\nu^2|xy|^{\nu-\frac{1}{2}}} \int_{0}^{\infty}\rmd k \,k\,\rme^{-\rmi \hbar t k^2/2m}\\[4mm]
    &&\times\left[{\rm J}_{\nu-\frac{1}{2}}(k|x|){\rm J}_{\nu-\frac{1}{2}}(k|y|)+ {\rm sgn\,}(xy){\rm J}_{\nu+\frac{1}{2}}(k|x|){\rm J}_{\nu+\frac{1}{2}}(k|y|)\right]
    \,.
\end{array}
\end{equation}
The remaining integration may be performed with the help of the integral formula 6.633/2 provided in table \cite{GR},
\begin{equation}\label{GR6633/2}
\int_{0}^{\infty}\rmd z\,z\,\rme^{-\alpha z^2} {\rm J}_\mu(az){\rm J}_\mu(bz)=\frac{1}{2\alpha}\exp\left\{-\frac{1}{4\alpha}(a^2 + b^2)\right\}\,{\rm I}_\mu\left(\frac{ab}{2\alpha}\right) \, ,
\end{equation}
which is valid for $a,b >0$, $\mu>-1$ and ${\rm Re\,}\alpha >0$. In order to accommodate the last condition we will assume that the mass $m$ has a small positive imaginary part, ${\rm Im\,}m >0$. Note that this is a common way to regularise Feynman path integrals \cite{Cameron1960,Cameron1962,Klauder2003}. The result then reads
\begin{equation}\label{KfreeI}
\begin{array}{rl}
  K_\nu(x,y;t)=&\displaystyle\frac{m}{\rmi\hbar t|xy|^{\nu-\frac{1}{2}}}\exp\left\{-\frac{m}{2\rmi\hbar t}(x^2+y^2 )\right\}\\[4mm]
  &\times \left[{\rm I}_{\nu-\frac{1}{2}}\left(\frac{m|xy|}{\rmi\hbar t}\right)+{\rm sgn\,}(xy) {\rm I}_{\nu+\frac{1}{2}}\left(\frac{m|xy|}{\rmi\hbar t}\right)\right]\,.
\end{array}\end{equation}
With the help of relation \eqref{Ereal} we can express the free particle propagator for WDQM in terms of the Dunkl kernel as follows
\begin{equation}\label{KfreeE}
  K_\nu(x,y;t)=\frac{(2\pi)^{\nu+\frac{1}{2}}}{c_\nu}
  \left(\frac{m}{2\pi\rmi \hbar t}\right)^{\nu+\frac{1}{2}}
  \rme^{\frac{\rmi }{\hbar }\frac{ m}{2t}(x^2+y^2)}\,E_\nu\left(\frac{mxy}{\rmi\hbar t}\right)\,.
\end{equation}
We note here, that for the undeformed case $\nu=0$ above propagator reduces to the well known result
\begin{equation}\label{K0free}
   K_0(x,y;t)=\sqrt{\frac{m}{2\pi\rmi \hbar t}}\, \rme^{\frac{\rmi }{\hbar }\frac{ m}{2t}(x -y)^2}\,.
\end{equation}

Having derived the free propagator in WDQM, we are now able to construct the path integral representation of the interacting system \eqref{H}. For this we will utilise the Lie-Trotter product formula,
\begin{equation}\label{LieTrotter}
  \rme^{-\frac{\rmi}{\hbar}H t}=\rme^{-\frac{\rmi}{\hbar}(H_\nu +V) t} = \lim_{N\to\infty} \left(\rme^{-\frac{\rmi}{\hbar}H_\nu t/N}\rme^{-\frac{\rmi}{\hbar}V t/N}\right)^N\,.
\end{equation}
Here we insert $N-1$ times the unity \eqref{unitop&ortho} in the form
\begin{equation}\label{unitiy}
  1= \int_{-\infty}^{+\infty}\rmd x_j\, |x_j|^{2\nu}|x_j\rangle\langle x_j|
\end{equation}
to arrive at
\begin{equation}\label{KPI}
\fl
\begin{array}{rrl}
  K(x,y;t)&:=&\langle x''|\rme^{-\frac{\rmi}{\hbar}H t}|x'\rangle\\
  &=&\displaystyle\lim_{N\to\infty}\prod_{j=1}^{N-1}\int_{-\infty}^{+\infty}\rmd x_j\, |x_j|^{2\nu}\prod_{j=1}^{N}K_\nu(x_j,x_{j-1};\varepsilon)\rme^{-(\rmi / \hbar)V(x_{j})\varepsilon},
\end{array}
\end{equation}
where we have set $\varepsilon:= t/N$, $x_N:=x$ and $x_0:=y$. Eq.\ \eqref{KPI} constitutes the path integral representation of WDQM propagator in its time-sliced notation.

In the standard path integral approach it suffices to consider terms up to order $\varepsilon$ within the path integral, where one may consider $\Delta x_j := x_j - x_{j-1}= O(\sqrt{\varepsilon})$ due to the Gaussian nature related to standard kinetic energy term, cf.\ \eqref{K0free}. See, for example, also the contribution of Inomata in part I of ref.\ \cite{Inomata1993}.
Hence, one may take the potential term at various "mid" points, $V(x_j)\varepsilon \simeq V(x_{j-1})\varepsilon \simeq V(\hat{x}_{j})\varepsilon$, with $\hat{x}_j^2:= x_jx_{j-1}$, all being equivalent within the path integral \eqref{KPI}.

Let us naively do the same approximation for the free short time propagator $K_\nu(x_j,x_{j-1};\varepsilon)$ appearing in \eqref{KPI}. That is, we utilise the first asymptotic relation in \eqref{Elargex} to get
\begin{equation}\label{Kshort}
  K_\nu(x_j,x_{j-1},\varepsilon) = \frac{1}{|\hat{x}_j|^{2\nu}}\sqrt{\frac{m}{2\pi\rmi\hbar\varepsilon}}
  \exp\left\{ \frac{\rmi }{\hbar }\left( \frac{m}{2\varepsilon}\Delta x_j^2-\frac{\nu^2}{m\hat{x}_j^2}\varepsilon\right) \right\}\,.
\end{equation}
This provides us with the path integral expression
\begin{equation}\label{KPI2}
\fl
\begin{array}{l}
  K(x,y;t)=\displaystyle\frac{1}{|xy|^\nu}\\
  \qquad\displaystyle \times
  \lim_{N\to\infty}
  \prod_{j=1}^{N-1}\int_{-\infty}^{+\infty}\rmd x_j\,
  \prod_{j=1}^{N}\sqrt{\frac{m}{2\pi\rmi\hbar\varepsilon}}\exp\left\{ \frac{\rmi }{\hbar }\left( \frac{m}{2\varepsilon}\Delta x_j^2-\frac{\nu^2}{m\hat{x}_j^2}\varepsilon-V(\hat{x}_j)\varepsilon\right) \right\}\,,
\end{array}
\end{equation}
Formally, the above expression may be written as
\begin{equation}\label{PIformal}
  K(x,y;t)=\frac{1}{|xy|^\nu} \int{\cal D}[x(s)]\,\rme^{(\rmi/\hbar)S[x(s)]}
\end{equation}
with action
\begin{equation}\label{S}
  S[x(s)]:= \int_{0}^{t}\rmd s\left[\frac{m}{2}\,\dot{x}^2(s)-V_{\rm eff}(x(s))\right]
\end{equation}
and effective potential
\begin{equation}\label{Veff}
  V_{\rm eff}(x):= \frac{\nu^2}{mx^2}+V(x)\,.
\end{equation}
Surprisingly, this expression is identical to the usual Feynman path integral in standard quantum mechanics. The only difference being the perfactor $\frac{1}{|xy|^\nu}$ and the additional repulsive singular potential $\frac{\nu^2}{mx^2}$ appearing in the effective potential \eqref{Veff}. Both indeed disappear in the undeformed limit $\nu\to 0$. However, recall that here we are working on a weighted Hilbert space characterise by parameter $\nu$. In addition the singular potential will also generate some subtleties. Therefore, we cannot blindly use the short-time approximation \eqref{Kshort} but have to keep in mind that there is a Dunkl kernel involved. This is similar to the situation in standard radial path integrals involving Bessel functions \cite{Inomata1993,Peak1969,Steiner1984,Fischer1992}.
As an explicit example let us discuss the harmonic oscillator.

The Dunkl Hamiltonian \eqref{H} with the harmonic oscillator potential $V(x)=\frac{m}{2}\omega^2x^2$, $\omega >0$, has been studied by many authors, see for example \cite{Rosenblum1994,Genest2013,Genest2014}. The corresponding spectral properties, that is, its eigenstates and eigenfunctions have been found in closed form. More recently, while studying the related coherent states \cite{Sedaghatnia2023}, the corresponding propagator was calculated by using its spectral representation and explicitly reads
\begin{equation}\label{KHO}
\fl
\begin{array}{rl}
  K^{\rm HO}(x,y;t)&:=\langle x|\exp\left\{-\frac{\rmi}{\hbar}\left(H_\nu +\frac{m}{2}\omega^2x^2\right)t\right\}|y\rangle\\
  &=\frac{1}{c_\nu}\left(\frac{m\omega}{\rmi\hbar\sin(\omega t)}\right)^{\nu+\frac{1}{2}}\exp\left\{\frac{\rmi m\omega}{2\hbar}\left({x}^2 + {y}^2\right)\cot(\omega t)\right\}
  {E}_\nu\left(\frac{m\omega xy}{\rmi \hbar \sin(\omega t)}\right)\,.
\end{array}
\end{equation}
The Feynman path integral expression \eqref{KPI} for this harmonic oscillator problem is then explicitly given by
\begin{equation}\label{KPIHO}
\fl
\begin{array}{l}
  K^{\rm HO}(x,y;t):=\displaystyle\lim_{N\to\infty}\prod_{j=1}^{N-1}\int_{-\infty}^{+\infty}\rmd x_j\, |x_j|^{2\nu}
  \prod_{j=1}^{N}K_\nu(x_j,x_{j-1};\varepsilon)\rme^{-\frac{\rmi m\omega^2}{4\hbar}(x^2_{j}+x^2_{j-1})\varepsilon}\,,
\end{array}
\end{equation}
where we have approximated the quadratic potential by its arithmetic mean value in the short time propagator
\begin{equation}\label{Khoshort}
\fl
  K_\nu(x_j,x_{j-1};\varepsilon)\rme^{-\frac{\rmi m\omega^2}{4\hbar}(x^2_{j}+x^2_{j-1})\varepsilon} =
  \frac{1}{c_\nu} \left(\frac{m}{\rmi \hbar \varepsilon}\right)^{\nu+\frac{1}{2}}
  \rme^{\frac{\rmi }{\hbar }\frac{ m}{2\varepsilon}(x_j^2+x_{j-1}^2)(1-\frac{\omega^2\varepsilon^2}{2})}\,{ E}_\nu\left(\frac{mx_jx_{j-1}}{\rmi\hbar \varepsilon}\right)\,.
\end{equation}
With following approximations within the path integral expression \eqref{KPIHO}
\begin{equation}\label{stappox}
  \varepsilon = \frac{\sin(\omega\varepsilon)}{\omega}\left[1+ O(\varepsilon^2)\right]\,,\qquad 1-\frac{\omega^2\varepsilon^2}{2}= \cos(\omega\varepsilon)\left[1+ O(\varepsilon^2)\right]
\end{equation}
we arrive at
\begin{equation}\label{KPIHO2}
\fl
\begin{array}{l}
  K^{\rm HO}(x,y;t):=\displaystyle\lim_{N\to\infty}\prod_{j=1}^{N-1}\int_{-\infty}^{+\infty}\rmd x_j\, |x_j|^{2\nu}\\
  \quad\times \displaystyle
  \prod_{j=1}^{N}\frac{1}{c_\nu}\left(\frac{m\omega}{\rmi\hbar\sin(\omega \varepsilon)}\right)^{\nu+\frac{1}{2}}\exp\left\{\frac{\rmi m\omega}{2\hbar}\left({x_j}^2 + {x_{j-1}}^2\right)\cot(\omega \varepsilon)\right\} {E}_\nu\left(\frac{x_jx_{j-1}m\omega}{\rmi \hbar \sin(\omega \varepsilon)}\right)\,.
\end{array}
\end{equation}
The expression within the product on the second line above is identical in form with the finite time propagator \eqref{KHO} with $t$ replaced by $\varepsilon=t/N$. Hence all $N-1$ integrals as well as the limit $N\to\infty$ are trivially performed resulting in the known finite-time propagator \eqref{KHO}.

This example shows that we may not naively use the asymptotic form \eqref{Kshort} of the Dunkl kernel but need to keep the Dunkl kernel as it is. What can be done is using approximations like \eqref{stappox} in its argument and the interacting potential $V(x)$. This is similar to what is known for a long time in standard radial path integrals where it is called asymptotic recombination technique \cite{Inomata1993}. The reason behind this is that the underlying stochastic process is not a Brownian motion but a Bessel process. In fact, the Euclidean path integral for the harmonic oscillator is in essence a well known solvable Bessel path integral \cite{Fischer1992}. This will be made more explicit when looking into the Feynman-Kac Formula for WDQM, which we will do in the following section.

\section{On the Feynman-Kac Formula for Wigner-Dunkl Quantum Mechanics}
Let us start by defining the Dunkl generator, which is basically given by the negative free Dunkl Hamiltonian \eqref{Hfree} in units such that $\hbar=m=1$,
\begin{equation}\label{DunklGen}
  {\cal L}^{(\nu)} := \frac{1}{2}\left[\frac{\partial^2}{\partial x^2} +\frac{2\nu}{x}\frac{\partial}{\partial x}-\frac{\nu}{x^2}(1-R)\right]\,.
\end{equation}
This operator is known to generate the so-called Dunkl process $X^{(\nu)}=({X}^{(\nu)}_\tau)_{\tau\geq 0}$ which is a stationary Markov process on the real line \cite{Roesler1998a,Roesler1998b}. The corresponding  transition probability density is given by
\begin{equation}\label{HKexplicit}
  d^{(\nu)}_\tau(x,y):= \langle x|\rme^{\tau {\cal L}^{(\nu)}}|y\rangle  = \frac{1}{c_\nu\tau^{\nu+\frac{1}{2}}} \, \,\rme^{-\frac{1}{2\tau}(|x|^2+|y|^2)}E_\nu\left(\frac{xy}{\tau}\right)\,,\quad \tau \geq 0\,.
\end{equation}
This in essence follows from the free particle propagator \eqref{KfreeE} when performing a  Wick rotation, where $\tau:= \rmi t >0$. Above transition density is the Euclidean propagator, also called Dunkl heat kernel, and is a well known result of Dunkl theory, see for example \cite{Roesler1998a}. The Dunkl transition density obeys the relations:
\begin{equation}\label{densityProperties}
\fl
\begin{array}{lll}
  (i) &{\rm Positivity:} & d^{(\nu)}_\tau(x,y) > 0\,, \\[2mm]
  (ii)&{\rm Initial~condition:} & \displaystyle \lim_{\tau\downarrow 0}d^{(\nu)}_\tau(x,y) = \delta(y-x)\,,\\[2mm]
  (iii)&{\rm Normalisation:} & \displaystyle\int_{-\infty}^{+\infty}\rmd x\,|x|^{2\nu}\, d^{(\nu)}_\tau(x,y) = 1\,,\\
  (iv)&{\rm Convolution:} & \displaystyle\int_{-\infty}^{+\infty}\rmd z\,|z|^{2\nu}\, d^{(\nu)}_{\tau_2}(x,z)d^{(\nu)}_{\tau_1}(z,y)=d^{(\nu)}_{\tau_1+\tau_2}(x,y)\,,
\end{array}
\end{equation}
and therefore fulfils Kolmogorov's extension theorem and constitutes a probability space together with a stochastic process on it, the Dunkl process \cite{Roesler1998b,Gallardo2005,Gallardo2006}. It must be noted that due to the presence of the non-local reflection operator $R$ in its generator \eqref{DunklGen}, this process is not a continuous process and exhibits jumps, it is a so-called c\`adl\`ag Markov process.\footnote{A c\`adl\`ag (French abbreviation for "continu à droite, limites à gauche") process has paths $\tau \mapsto X^{(\nu)}_\tau$ which are right-continuous with a left limit.} We also note that in the above the measure of our weighted Hilbert space plays the role of the speed measure $m_{{\rm D}^{(\nu)}}(x)\rmd x= |x|^{2\nu} \rmd x$ for the Dunkl process.

Let ${\cal C}(\mathbb{R},y)$ denote the space of c\`adl\`ag paths $X^{(\nu)}_\tau$ starting at $\tau =0$ in $y\in\mathbb{R}$. Let $V$ be bounded from below and symmetric, i.e.\ $V(x)=V(-x)=(RV)(x)$. Then the Feynman-Kac formula for the Dunkl process reads
\begin{equation}\label{FKDunkl}\fl
  \langle x|\rme^{\tau \left({\cal L}^{(\nu)} -V\right)}|y\rangle = \int_{{\cal C}(\mathbb{R},y)}\rmd{\bf D}^{(\nu)}[z(s)]\,\delta(z(\tau)-x'')\, \exp\left\{-\int_{0}^{\tau}\rmd s\,V(x(s))\right\}\,,
\end{equation}
where $\rmd{\bf D}^{(\nu)}$ denotes the measure on ${\cal C}(\mathbb{R},y)$ induced by the transition density \eqref{HKexplicit}.

We now will investigate this expression and try to reduce it to the well-known Bessel processes. In doing so we start with introducing a pair of projection operators
\begin{equation}\label{Projectors}
  P_\pm :=\frac{1}{2}\left(1\pm R\right)\,,
\end{equation}
which obviously obey the relations
\begin{equation}\label{Pprop}
  P_\pm^2 =P_\pm\,,\qquad P_+P_- = 0= P_-P_+\,,\qquad P_+ + P_- =1\,,
\end{equation}
and thus form a complete set of ortho-normal projectors. That is, we can decompose our Hilbert space into two subspaces
\begin{equation}\label{Hpm}
  {\cal H} = {\cal H}_+ \oplus {\cal H}_-\qquad\mbox{with}\qquad {\cal H}_\pm := P_\pm {\cal H} P_\pm\,.
\end{equation}
That is ${\cal H}_\pm$ is the eigenspace of $R$ associated with its eigenvalues $\pm 1$, respectively. With $f_\pm(x):= (P_\pm f)(x)$ we can decompose any function $f\in{\cal H}$ into its even and odd part, $f(x)=f_+(x)+f_-(x)$. This also applies to operators and in particular to the generator
\begin{equation}\label{DunklGen+-0}
  {\cal L}^{(\nu)}_\pm := P_\pm {\cal L}^{(\nu)} P_\pm\,.
\end{equation}
Being a bit more explicit, the even and odd parts of the Dunkl generator read
\begin{equation}\label{DunklGen+-}
  {\cal L}^{(\nu)}_+ = \frac{1}{2}\left[\frac{\partial^2}{\partial x^2} +\frac{2\nu}{x}\frac{\partial}{\partial x}\right]\,,
  \qquad
  {\cal L}^{(\nu)}_- = {\cal L}^{(\nu)}_+ -\frac{\nu}{x^2}\,.
\end{equation}
We note that for symmetric potential $V$ the reflection operator commutes with it and hence we have the relation $V=P_\pm V P_{\pm}$. We also note the relations $P_+|x\rangle = \left|\, |x|\right\rangle $ and $P_-|x\rangle = {\rm sgn\,}x\,\left|\, |x|\right\rangle $. Having this in mind we are able to rewrite the left-hand side of \eqref{FKDunkl} as follows
\begin{equation}\label{DunklHeatKernel}\fl
  \left\langle x\left|\rme^{\tau \left({\cal L}^{(\nu)} -V\right)}\right|y\right\rangle = \left\langle |x|\,\left|\rme^{\tau \left({\cal L}_+^{(\nu)} - V\right)}\right|\,|y|\right\rangle
   +{\rm sgn\,}(xy)\,\left\langle |x|\,\left|\rme^{\tau \left({\cal L}_-^{(\nu)} - V\right)}\right|\,|y|\right\rangle \,.
\end{equation}
Both kernels on the right-hand-side above live now on the positive half line and hence we may basically replace the generators ${\cal L}^{(\nu)}_\pm$ acting on ${\cal H}_\pm$ by the Bessel generators ${\cal L}_{\rm B}^{(\nu\mp\frac{1}{2})}$, see eq.\ \eqref{BesselGen} in appendix A. Here ${\cal L}_{\rm B}^{(\nu+\frac{1}{2})}$ acts on functions $f:\mathbb{R}^+\to\mathbb{R}$ with $f(0)=0$, i.e. Dirichlet boundary condition are imposed. However, for ${\cal L}_{\rm B}^{(\nu-\frac{1}{2})}$ a Neumann condition must be imposed, i.e.\ $f'(0) = 0$.

At this stage we remark that both generators \eqref{DunklGen+-} are in essence generators of a Bessel process with index $\alpha = \nu-\frac{1}{2}$ and  $\beta = \nu+\frac{1}{2}$, respectively. Here we refer to appendix A for more details.
We also remark that a simple calculation using the explicit expressions \eqref{Ereal}, \eqref{HKexplicit} and \eqref{Bdensity} provides us with a relation between the Dunkl density and the Bessel densities as follows
\begin{equation}\label{denstiyrelation}
  d_\tau^{(\nu)}(x,y) = b_\tau^{(\nu-\frac{1}{2})}(|x|,|y|) + x y\,b_\tau^{(\nu+\frac{1}{2})}(|x|,|y|)\,,
\end{equation}
Hence, we finally arrive at the Feynman-Kac formula
\begin{equation}\label{FKDunkl2}
\fl
\begin{array}{l}
\langle x|\rme^{\tau \left({\cal L}^{(\nu)} -V\right)}|y\rangle = \\[4mm]
\displaystyle\qquad
\int_{{\cal C}(\mathbb{R}^+,|y|)}\rmd{\bf B}^{(\nu-\frac{1}{2})}[z(s)]\,\delta(z(\tau)-|x|)\, \exp\left\{-\int_{0}^{\tau}\rmd s\,V(z(s))\right\}\\[4mm]
\qquad\displaystyle +\, xy
\int_{{\cal C}(\mathbb{R}^+,|y|)}\rmd{\bf B}^{(\nu+\frac{1}{2})}[z(s)]\,\delta(z(\tau)-|x|)\, \exp\left\{-\int_{0}^{\tau}\rmd s\,V(z(s))\right\}\,.
\end{array}
\end{equation}
We note that this is consistent with relation \eqref{DunklHeatKernel} together with the change of index as discussed in the appendix, cf.\ eq.\ \eqref{BesselFK3}.
In this way we have reduced the Dunkl-type path integral \eqref{FKDunkl} to the sum of two well-studied Bessel-type path integrals. Whereas the Dunkl process has jumps, the two Bessel processes appearing on the right-hand side of \eqref{FKDunkl2} have continues paths. As a side remark we note that for the special case $\nu =0$ relation \eqref{denstiyrelation} results in Wiener's density on the real line as expected
\begin{equation}\label{nu=0}
   d_\tau^{(0)}(x,y) = \frac{1}{\sqrt{2\pi\tau}}\,\exp\left\{-\frac{(x-y)^2}{2\tau}\right\}\,.
\end{equation}

As an explicit example we choose, as in the previous section, the harmonic oscillator potential $V(x)=\frac{1}{2}\omega^2x^2$. Both Bessel path integrals on the right-hand side of \eqref{FKDunkl2} can explicitly be calculated, see eq.\ \eqref{BesselFKHO} in the appendix. A little calculation brings us to the result
\begin{equation}\label{FKHO}
\fl
\langle x|\rme^{\tau \left({\cal L}^{(\nu)} -\frac{1}{2}\omega^2x^2\right)}|y\rangle = \frac{1}{c_\nu}\left(\frac{\omega}{\sinh(\omega\tau)}\right)^{\nu+\frac{1}{2}}
\,\rme^{-\frac{1}{2\tau}(x^2+y^2)\coth(\omega\tau)}E_\nu\left(\frac{\omega xy}{\sinh(\omega\tau)}\right)\,,
\end{equation}
which is the Euclidean version of \eqref{KHO}. See, for example, also ref.\ \cite{Amri2019}.

\section{Summary and Outlook}
The objective of the present work was to study the time evolution, in both real and Euclidean time, of Wigner-Dunkl quantum mechanics. We could show that the free time evolution of a Gaussian wave packet exhibits the same dispersion known from ordinary quantum mechanics. We calculated the propagator of the free WDQM and utilised this together with the Lie-Trotter formula to arrive at the Feynman path integral expression \eqref{KPI}. By turning to Euclidean time we could show that the underlying stochastic process is represented by a reflecting and an absorbing Bessel process. An explicit form of the associated transition density was derived in \eqref{denstiyrelation}, which finally led us to the Dunkl-Feynman-Kac formula \eqref{FKDunkl2}. The path integral for the harmonic interaction was explicitly solved.

The current approach may also be used to solve the path integral for other interactions. For example, the singular potential $g/x^2$ may be absorbed via a change of index in the Bessel processes \cite{Fischer1992}. It may even be possible to tackle the one-dimensional hydrogen atom problem, which, via path–dependent time transformations, may be mapped on a Bessel process with harmonic interaction \cite{Fischer1995}.  This is of particular interest as we may establish a Newton-Hook-like duality or more generally a power-law duality also within the WDQM formalism. See e.g.\ the recent work by Inomata et al \cite{Inomata2024}.

\appendix
\section{Short summary on Bessel processes}
This appendix collects some basic facts about Bessel processes \cite{Revuz1999} with index $\alpha > -1$. The generator of the Bessel process ${\bf B}^{(\alpha)}$ is defined by
\begin{equation}\label{BesselGen}
  {\cal L}_{\rm B}^{(\alpha)} := \frac{1}{2}\left[\frac{\partial^2}{\partial x^2} +\frac{2\alpha +1 }{x}\frac{\partial}{\partial x}\right]\,,\qquad\alpha > -1\,,
\end{equation}
with state space given by the positive half line $\mathbb{R}^+:=\{x\in\mathbb{R} | x\geq 0\}$.
For $d:=2\alpha +2>0$ it is also called $d$-dimensional Bessel process as for positive integer $d$ it represents the radial part of the Wiener process in $d$ dimensions. With the speed measure
$m_{{\rm B}^{(\alpha)}}(x)\rmd x:=2x^{2\alpha +1}\rmd x$, see e.g.\ \cite{Getoor1979,Shimizu2017,Serafin2017}, the corresponding transition density is given in  a symmetric form
\begin{equation}\label{Bdensity}
  b^{(\alpha)}_\tau(x,y):= \frac{1}{2\tau}\left(xy\right)^{-\alpha}\rme^{-\frac{1}{2\tau}(x^2+y^2)}{\rm I}_{\alpha}\left(\frac{xy}{\tau}\right)\,,\qquad x,y\geq 0\,,
\end{equation}
and obeys the convolution relation
\begin{equation}\label{Bconvolution}
  b^{(\alpha)}_{\tau_1+\tau_2}(z,x)= \int_{0}^{\infty}\rmd y\, m_{{\rm B}^{(\alpha)}}(y)\, b^{(\alpha)}_{\tau_2}(z,y)b^{(\alpha)}_{\tau_1}(y,x)\,.
\end{equation}
Obviously, here we are dealing with a weighted Hilbert space $L^2\left(\mathbb{R},m_{{\rm B}^{(\alpha)}}(x)\rmd x\right)$.
In the literature one often finds the asymmetric transition density with the speed measure being absorbed in it
\begin{equation}\label{altBdensity}
  \tilde{b}^{(\alpha)}_\tau(x,y):= m_{{\rm B}^{(\alpha)}}(x)\,b^{(\alpha)}_\tau(x,y)= \frac{x}{\tau}\left(\frac{x}{y}\right)^{\alpha}\rme^{-\frac{1}{2\tau}(x^2+y^2)}\, {\rm I}_{\alpha}\left(\frac{xy}{\tau}\right)\,.
\end{equation}
Here the generator \eqref{BesselGen} acts on states in the unweighed Hilbert space $L^2(\mathbb{R},\rmd x)$.

The Feynman-Kac formula for Bessel processes on the positive real line reads
\begin{equation}\label{BesselFK}
\fl
  \langle x|\rme^{\tau \left({\cal L}_{\rm B}^{(\alpha)}-V \right)}|y\rangle = \int_{{\cal C}(\mathbb{R}^+,y)}\rmd{\bf B}^{(\alpha)}[z(s)]\,\delta(z(\tau)-x)\, \exp\left\{-\int_{0}^{\tau}\rmd s\,V(z(s))\right\}\,.
\end{equation}
Two Bessel processes with different index, say $\alpha$ and $\beta$, are related to each other by the so-called Radon-Nikodym derivative, which results in the relation
\begin{equation}\label{BesselFK2}
\fl
\begin{array}{l}
\displaystyle
  \int_{{\cal C}(\mathbb{R}^+,y)}\rmd{\bf B}^{(\alpha)}[z(s)]\,\delta(z(\tau)-x)\, \exp\left\{-\int_{0}^{\tau}\rmd s\,V(z(s))\right\}=\\[4mm]
  \displaystyle\quad \left(xy\right)^{\beta- \alpha}
  \int_{{\cal C}(\mathbb{R}^+,y)}\rmd{\bf B}^{(\beta)}[x]\,\delta(z(\tau)-x)\, \exp\left\{-\int_{0}^{\tau}\rmd s\,\left[V(z(s))+ \frac{\alpha^2 - \beta^2}{2z^2(s)}\right]\right\}\,.
\end{array}\end{equation}
Here we note that in the literature, by using the asymmetric Bessel density \eqref{altBdensity} one finds as prefactor on the right-hand side the expression
$\left(\frac{x}{y}\right)^{\alpha - \beta}$. In our case, when using the symmetric density \eqref{Bdensity} we need to correct this with the speed densities, that is,
$ \frac{m_{{\rm B}^{(\beta)}}(x)}{m_{{\rm B}^{(\alpha)}}(x)}\,\left(\frac{x}{y}\right)^{\alpha - \beta} =(xy)^{\beta -\alpha} $.
In particular for $\alpha = \nu-\frac{1}{2}$ and $\beta = \nu+ \frac{1}{2}$ this reads
\begin{equation}\label{BesselFK3}
\fl
\begin{array}{l}
  \langle x|\exp\left\{\tau \left({\cal L}_{\rm B}^{(\nu - \frac{1}{2})}-\frac{\nu}{x^2}-V \right)\right\}|y\rangle
  =\displaystyle xy\,\langle x|\exp\left\{\tau \left({\cal L}_{\rm B}^{(\nu + \frac{1}{2})}-V \right)\right\}|y\rangle\\[4mm]
  \qquad=\displaystyle xy\int_{{\cal C}(\mathbb{R}^+,y)}\rmd{\bf B}^{(\nu + \frac{1}{2})}[z(s)]\,\delta(z(\tau)-x)\, \exp\left\{-\int_{0}^{\tau}\rmd s\,V(z(s))\right\}\,.
\end{array}\end{equation}
We also recall the well-known path-integral relation for the radial harmonic oscillator where $V(z)=\frac{1}{2}\omega^2z^2$, see e.g.\ \cite{Fischer1992},
\begin{equation}\label{BesselFKHO}
\fl
\begin{array}{l}
  \langle x|\rme^{\tau \left({\cal L}_{\rm B}^{(\alpha)}-\frac{1}{2}\omega^2z^2 \right)}|y\rangle
  \displaystyle= \int_{{\cal C}(\mathbb{R}^+,y)}\rmd{\bf B}^{(\alpha)}[z(s)]\,\delta(z(\tau)-x'')\, \exp\left\{-\frac{\omega^2}{2}\int_{0}^{\tau}\rmd s\,z^2(s)\right\}\\
  \qquad\displaystyle= \frac{1}{{2}(xy)^\alpha}\frac{\omega}{\sinh(\omega\tau)}\exp\left\{-\frac{\omega}{2}(x^2+y^2)\coth(\omega\tau) \right\}{\rm I}_\alpha\left(\frac{\omega xy}{\sinh(\omega\tau)}\right)
  \,.
\end{array}
\end{equation}
Again note that here we have a symmetric expression as compared to ref.\ \cite{Fischer1992}, which is due to the fact that we have the speed measure as weight in our Hilbert space.

\end{document}